\documentclass[letterpaper, 10pt, journal]{IEEEtran}  
\IEEEoverridecommandlockouts 
\usepackage{mathtools} 
\usepackage{hyperref, graphicx} 
\usepackage{amsmath} 
\usepackage{amssymb}  
\usepackage{cite}
\usepackage{multirow}
\usepackage{color}
\usepackage{algorithm, algorithmicx, algpseudocode}

\newcommand{\R}{{\mathbb R}}

\newcommand{\Rnn}{{\mathbb R}_{\ge 0}}
\newcommand{\Rp}{{\mathbb R}_{> 0}}
\newcommand{\C}{{\mathbb C}}

\newcommand{\cA}{{\mathcal A}}

\newcommand{\cB}{{\mathcal B}}
\newcommand{\cC}{{\mathcal C}}
\newcommand{\cD}{{\mathcal D}}
\newcommand{\cE}{{\mathcal E}}

\newcommand{\cP}{{\mathcal P}}
\newcommand{\cQ}{{\mathcal Q}}

\newcommand{\cS}{{\mathcal S}}
\newcommand{\cT}{{\mathcal T}}
\newcommand{\cU}{{\mathcal U}}

\newcommand{\Lone}{{\mathbb L_1}}

\newcommand{\diag}{{\mathrm{diag}}}

\newcommand{\argmin}{\mathop{\mathrm{argmin}}}

\newcommand{\dom}{\mathrm{dom}}

\def\QED{\mbox{\rule[0pt]{1.3ex}{1.3ex}}} 

\newenvironment{proof-of}[1]{{\it Proof of #1:\,}}{\hfill\QED\par}

\newtheorem{defn}{Definition}

\newtheorem{prop}{Proposition}

\title{Pulse-Based Control Using {K}oopman Operator Under Parametric Uncertainty
\author{Aivar Sootla and Damien Ernst
		\thanks{Aivar Sootla is with the Department of Engineering Science, University of Oxford, Parks Road, Oxford, OX1 3PJ, UK {\tt aivar.sootla@eng.ox.ac.uk}. Damien Ernst is with the Montefiore Institute, University of Li\`{e}ge, B-4000, Belgium {\tt dernst@ulg.ac.be}.}
		\thanks{ This work was performed while A. Sootla held a postdoctoral position at University of Li\`{e}ge funded by F.R.S.--FNRS. Currently, A. Sootla is supported by the EPSRC Grant EP/M002454/1. 
		}
	}
}
\begin{document}
\maketitle

\begin{abstract}
In applications, such as biomedicine and systems/synthetic biology, technical limitations in actuation complicate implementation of time-varying control signals. In order to alleviate some of these limitations, it may be desirable to derive simple control policies, such as step functions with fixed magnitude and length (or temporal pulses). In this technical note, we further develop a recently proposed pulse-based solution to the convergence problem, i.e., minimizing the convergence time to the target exponentially stable equilibrium, for monotone systems. In particular, we extend this solution to monotone systems with parametric uncertainty. Our solutions also provide worst-case estimates on convergence times. Furthermore, we indicate how our tools can be used for a class of non-monotone systems, and more importantly how these tools can be extended to other control problems. We illustrate our approach on switching under parametric uncertainty and regulation around a saddle point problems in a genetic toggle switch system. 
\end{abstract}


\section{Introduction} 
Recent developments in molecular biology, neuroscience, bioengineering among others, opened control theory to new fields of applications. If the implementation of optimal control methods in biological and biomedical applications is technically feasible, it may still be costly, labor intensive and/or invasive. This drawback in part stems from actuating using a time-varying feedback control signal, which is typically computed using an optimal control framework such as model predictive control, for example. Hence it may be desirable to use (piecewise) constant control signals in order to simplify the actuation. We motivate our formulation through concrete limitations and constraints by considering synthetic biology as a motivating application, but our results can be applied to a broader class of problems. 

Synthetic biology aims to engineer and control biological functions in living cells and is an emerging field of science with numerous applications~\cite{Purnick:2009}. Recently, control theoretic regulation of protein levels in microbes was shown to be possible (cf.~\cite{milias2011silico, uhlendorf2012long}). One of the simplest, but nevertheless important, control problems is a switching problem. For example, it may be desirable to switch off production of a certain protein in a cell by an external stimulus (by adding a chemical, heat shock, exposure to light etc). Switching between two modes (production of a protein is ``on'' and ``off'') can mathematically be expressed as switching between stable equilibria of a model. This problem can be solved using standard optimal control tools, however, in general, actuation has some limitations. If, for instance, we use the addition of a chemical as an actuation technique, then lowering the chemical concentration in a culture can be done through dilution, but it is more labor intensive than increasing the concentration. It may, therefore, be desirable to derive (piecewise-)constant (or at least easy-to-implement) control policies, which can also solve the problem. 

In~\cite{sootla2015pulsesaut}, \cite{sootla2016nolcos}, it has been proposed to solve \emph{the switching problem} using temporal pulses of a fixed length $\tau$ and a fixed magnitude $\mu$ for monotone systems (cf.~\cite{angeli2003monotone}). Since monotone systems are met often in biological applications~\cite{sontag2007monotone}, restricting the class of system to monotone is not unreasonable. In~\cite{sootla2016optimalswitching}, the authors extended this framework by solving \emph{a convergence problem} using temporal pulses, i.e. driving the system from an arbitrary initial point to the target exponentially stable equilibrium. This allowed a closed-loop formulation of the pulse-based control, thus providing robustness through feedback. Furthermore, the problem was reduced to a static optimization problem using the Koopman operator framework (cf.~\cite{mezic2005}). In this technical note, we further develop the results of~\cite{sootla2015pulsesaut} and~\cite{sootla2016optimalswitching}. In particular, we extend the framework to control under parametric uncertainty using the Koopman operator and monotone system theory. We also use our framework as a building block to solve other control problems, for example, we apply our method to the problem of regulation around a saddle point. 

The rest of the note is organized as follows. In Section~\ref{s:prel}, we cover the essential background on monotone systems and Koopman operator. In Section~\ref{s:switching}, we discuss how to solve the switching problem in a closed loop setting using the Koopman operator, while generalizing and extending the problem formulation. In Section~\ref{ss:param-unc}, we consider the convergence problem under parametric uncertainty and illustrate the framework on numerical examples. The proofs of our technical results are found in Appendix~\ref{app:proofs}.

\section{Preliminaries}
\label{s:prel}
We consider a system in the following form
\begin{equation}
\label{sys:f}
\dot x = f(x,p, u),\quad x(0) = x_0,
\end{equation} 
where $f: \cD\times \cP \times \cU\rightarrow \R^n$,  $u:\R_{\ge 0}\rightarrow \cU$, where $\cD\subseteq\R^n$, $\cP \subseteq\R^k$, and $u$ belongs to the space $\cU_{\infty}$ of Lebesgue measurable functions with values from $\cU\subset\R$ (i.e., the system is single-input). We define the flow map $\phi: \R \times \cD \times \cP \times \cU_{\infty}\rightarrow \R^n$, where $\phi(t, x_0, p, u)$ is a solution to the system~\eqref{sys:f} with an initial condition $x_0$, parameter value $p$ and a control signal $u$. If $u=0$, then we call the system~\eqref{sys:f} \emph{unforced}. Throughout the note, we assume that $f\in C^2$ and we denote the Jacobian matrix of $f(x,p,0)$ as $J(x,p)$. For every $p$, we assume that the eigenvectors of $J(x^\ast(p),p)$ are linearly independent (i.e., $J(x^\ast(p), p)$ are diagonalizable), where $x^\ast(p)$ is a stable equilibrium of $\dot x = f(x, p, 0)$. We denote the eigenvalues of $J(x^\ast(p), p)$ by $\lambda_i(p)$, we enumerate them counting their algebraic multiplicities, i.e., $i = 1,\dots, n$. 
We also assume that the eigenvalues $\lambda_j(p)$ are ordered according to their real parts, i.e.,  $\Re(\lambda_i(p)) \ge \Re(\lambda_j(p))$, for $i\le j$. If $\lambda_1(p)$ is such that $\Re(\lambda_1(p))> \Re(\lambda_2(p))$, then $\lambda_1(p)$ is called \emph{the dominant eigenvalue}. We will drop the dependence on $p$ and write $x^\ast$, $J(x^\ast)$, $\lambda_i$ when the parameter value is clear from the context. 

\emph{Koopman Operator}.  Spectral properties of nonlinear dynamical systems $\dot x = f(x)$ can be described through the Koopman operator. We limit our study of the Koopman operator to systems with $f\in C^2$ and admitting an exponentially stable equilibrium $x^\ast$, i.e., $J(x^\ast)$ is a Hurwitz matrix (its eigenvalues are such that $\Re(\lambda_j) <0$ for all $j$). We assume that $\cB(x^\ast)$ is the basin of attraction of $x^\ast$, that is, $\cB(x^\ast) = \left\{ x\in \R^n \bigl| \lim_{t\rightarrow\infty}\phi(t,x) = x^\ast \right\}$.

The Koopman semigroup of operators associated with $\dot x = f(x)$ acts on functions $g:\R^n\to \C$ (also called observables) and is defined as 
\begin{gather}
U^t g(x) = g(\phi(t,x)),
\end{gather}
 where $\phi(t,x)$ is a flow of the system, and $x\in\cB(x^\ast)$. The Koopman semigroup is linear~\cite{mezic2005}, and we can define the Koopman eigenfunctions as nontrivial functions $s:\cB(x^\ast)\rightarrow \C$ satisfying $U^t s(x) = s(\phi(t, x)) = s(x) \, e^{\lambda t}$ for all $t>0$, where $e^{\lambda t}$ belong to the point spectrum of $U^t$ and $\lambda$'s are called Koopman eigenvalues. Under the assumptions above, it can be shown that the eigenvalues $\lambda_j$ of the Jacobian matrix $J(x^\ast)$ are also Koopman eigenvalues (cf.~\cite{mauroy2014global}). Furthermore, there exist $n$ eigenfunctions $s_j \in C^1(\cB(x^\ast))$ associated with eigenvalues $\lambda_j$~\cite{mauroy2014global} and
\begin{gather}
(f(x))^T \nabla s_j(x) = \lambda_j s_j(x). \label{eq:s-one}
\end{gather}
The eigenfunction $s_1(x)$ can be computed at a given point using  \emph{Laplace averages}~\cite{mauroy2013isostables}. Furthermore, the eigenfunctions $s_j(x)$  can be estimated using linear algebraic methods~\cite{mauroy2014global}. An important concept related to the eigenfunction $s_1$, which is associated with the Koopman eigenvalue $\lambda_1$, is called isostables.

\begin{defn} Suppose $s_1(x)$ is a $C^1(\cB(x^\ast))$ eigenfunction corresponding to the dominant eigenvalue $\lambda_1$ with $0> \Re(\lambda_1) > \Re(\lambda_j)$ for $j\ge 2$. The \emph{isostables} $\partial \cB_\alpha$ for $\alpha\ge 0$ are boundaries of the sublevel sets $s_1$ (i.e., $\cB_\alpha =  \{x \in \cB(x^\ast) | |s_1(x)| \le \alpha \}$), or $\partial\cB_\alpha=\left\{x \in \cB(x^\ast)\Bigl| |s_1(x)|=\alpha\right\}$. 
We will also use the following notation $\partial_+ \cB_\alpha=\left\{x\in \cB(x^\ast)\Bigl| s_1(x) = \alpha\right\}$ and $\partial_- \cB_\alpha=\left\{x\in \cB(x^\ast)\Bigl| s_1(x) = -\alpha\right\}$ for positive $\alpha$.
\end{defn}

It can be also shown that the eigenfunction $s_1(x)$ captures the dominant behavior of the system, and $\cB_\alpha\rightarrow\cB$ as $\alpha\rightarrow\infty$. Furthermore, trajectories with initial conditions on the same isostable $\partial \cB_{\alpha_1}$ reach other isostables $\partial \cB_{\alpha_2}$ after a time
\begin{equation}
\label{time_isostable}
\cT = \frac{1}{|\Re(\lambda_1)|} \ln \left(\frac{\alpha_1}{\alpha_2}\right), \text{ if } \alpha_2 < \alpha_1
\end{equation}

For an eigenvalue $\lambda_1$ with multiplicity $\mu_1$, there exist $\mu_1$ eigenfunctions, and hence the isostables are not unique. However, we will consider specific eigenfunctions and isostables in the case of monotone systems. For more information about the isostables see~\cite{mauroy2013isostables}. We will also use the following concept introduced in~\cite{sootla2016optimalswitching}.

\begin{defn} \label{def:control-smth-function} 
	Let $r:\R^n\times \Rnn\times \Rnn \rightarrow \C\bigcup\{\infty\}$ be called a \emph{pulse control function} and defined as:
	\begin{gather*}
	r(x, \mu,\tau) = s_1(\phi(\tau, x, \mu)), 
	\end{gather*}
	where $s_1$ is a dominant eigenfunction on $\cB(x^\ast)$, and we use the convention that $r(x, \mu, \tau) = \infty$, if $\phi(\tau, x, \mu) \not \in \cB(x^\ast)$.	
\end{defn}

\emph{Monotone Systems and their Spectral Properties.} We define a partial order $\succeq$ as follows: $x\succeq y$ if and only if $ x - y \in \Rnn^n$.  We will also write $x\succ y$ if $x\succeq y$ and $x\ne y$, and $x\gg y$ if $x- y \in \Rp^n$. We define a partial order on the space of signals $u\in \cU_{\infty}$: $u\succeq v$ if $u(t) - v(t) \in \Rnn$ for all $t\ge 0$. Let $[x,~y]$ denote an \emph{order-interval}, that is $[x,~y] = \{ z\in\R^n | x\preceq  z \preceq y \}$. We call a set $\cA$ \emph{order-convex} if for all $x$, $y$ in $\cA$ the interval $[x,~y]$ is a subset of $\cA$. We note that, in general, partial orders can be induced by other cones in $\R^n$ (cf.~\cite{smith2008monotone}).

\begin{defn} Consider a function $W:\R^n \rightarrow \R\cup \{-\infty, +\infty\}$.
	 The effective domain of $W$ is defined as $\dom(W) = \{x \in\R^n | |W(x)| < \infty  \}$. We call $W$ \emph{increasing}  if $W(x) \le W(y)$ for all $x\preceq y$ on $\dom(W)$. 
\end{defn}
\begin{defn}\label{def:mon}
The system~\eqref{sys:f} is called \emph{monotone} if $\phi(t,x, p, u)\preceq \phi(t,y, q, v)$ for all $t\ge 0$, and for all $x\preceq y$, $p\preceq q$, and $u\preceq v$. 
\end{defn}

A certificate for monotonicity involves conditions on the vector field of the system and is usually called	
\emph{Kamke-M\"uller} conditions~(cf.~\cite{angeli2003monotone}). We finally mention spectral properties of monotone systems (\cite{sootla2016geometry}).
\begin{prop} \label{prop:mon-eig-fun}
	Consider the system $\dot x = f(x)$ with $f\in C^2$, which admits an exponentially stable equilibrium $x^\ast$ with a basin of attraction $\cB(x^\ast)$. Assume that $J(x^\ast)$ is diagonalizable with eigenvalues $\lambda_j$ such that $\Re(\lambda_1) > \Re(\lambda_j)$ for all $j\ge 2$. If the system is monotone, then $\lambda_1$ is real and negative, $v_1$, which is a right eigenvector of $J(x^\ast)$ corresponding to $\lambda_1$, and $s_1$, which is a Koopman eigenfunction corresponding to $\lambda_1$, can be chosen such that $v_1 \succeq 0$, $\nabla s_1(x) \succeq 0$ for all $x\in \cB(x^\ast)$ and $v_1^T \nabla s_1(x^\ast) = 1$. 
\end{prop}

Without loss of generality, we assume that for monotone systems $s_1(x)$ is such that $\nabla s_1(x)\succeq 0$ for all $x\in\cB(x^\ast)$. 

\section{Pulse-Based Control Using Koopman Operator} \label{s:switching}
\subsection{Convergence to an Isostable} \label{ss:sol-koopman}
The main motivating problem in~\cite{sootla2016optimalswitching} is a convergence problem from an arbitrary initial point $x^0$ to an $\varepsilon$-ball around the target equilibrium $x^\bullet$  (e.g., in the Euclidean metric). This problem can be solved using standard optimal control methods, however, the optimal control policy can have a complicated dependence on time (in the case of the open-loop control) or state (in the case of the closed-loop control). Therefore the authors proceeded in a different direction and formulated another problem as follows.

\emph{Problem 1: Converging to an isostable.} Consider the system $\dot x = f(x,u)$ at the initial state $x^0$ and control signals 
\begin{equation}
u(t) = \mu h(t,\tau), \textrm{ where  } 
h(t,\tau) = 
\begin{cases} 1 & 0 \leq t \leq \tau\,,\\
0 & t >\tau\,.
\end{cases} \label{eq:pulse}
\end{equation}
Compute $\mu$, $\tau$ such that the flow $\phi(t,x^0, \mu h(\cdot, \tau))$ reaches $\cB_\varepsilon(x^\bullet)$ with $\varepsilon\ll 1$ in minimum time $\cT_{\rm conv}$ subject to an energy budget $\|u\|_\Lone =\mu\cdot \tau\le \cE_{\rm max}$.

The authors argued that if $\cT_{\rm conv}$, $\tau$, $\mu$ obtained from Problem 1 are such that $\cT_{\rm conv}\gg \tau$, or $\mu$ is sufficiently small, then we get a good approximation of the convergence time to an $\varepsilon$-ball around $x^\bullet$. In order to simplify the problem the following assumptions are made:
\begin{enumerate}
	\item[{\bf A1.}] Let $f(x,u)$ be $C^2$ in $(x,u)$ on $\cD\times \cU$.
	\item[{\bf A2.}] Let the system $\dot x = f(x, 0)$ have an exponentially stable equilibrium $x^\bullet$ with a basin of attraction $\cB(x^\bullet)\subseteq\cD$.
	\item[{\bf A3.}] Let the system be monotone on $\cD\times\cU$. 
	\item[{\bf A4.}] The dominant eigenfunction $s_1(x)$ of the system $\dot x = f(x,0)$ defined on the basin of attraction $\cB(x^\bullet)$ is such that  $\nabla s_1(x) \gg 0$ for all $x\in \cB(x^\bullet)$. 
	\item[{\bf A5.}] $f(x, \mu_1) \succ f(x, \mu_2)$ for all $x\in\cD$ and $\mu_1> \mu_2\ge 0$.
\end{enumerate}

Assumption~{A1} guarantees existence and uniqueness of solutions of~$\dot x =f(x, u)$ and existence of $C^1$ eigenfunctions around an exponentially stable equilibrium. We ensure that the target equilibrium is exponentially stable in Assumption~A2. Note that in the case of exponentially stable monotone systems with $f\in C^2$ (Assumptions~{A1}-{A3}), we have that $\nabla s_1(x) \succeq 0$ and $f(x, \mu_1) \succeq f(x, \mu_2)$ for all $x\in\cB(x^\bullet)$, $\mu_1 \ge \mu_2$ (see~\cite{angeli2003monotone}). Therefore Assumption~A4 holds if, for example, the Jacobian of $f(x, 0)$ is irreducible for all $x$, while Assumption~A5 additionally requires that $f(x, \mu_1) \ne f(x, \mu_2)$ for all $x\in\cB(x^\bullet)$, $\mu_1 > \mu_2$. Therefore by making Assumptions~A4 and~A5 we do not add significant restrictions in comparison to the previous assumptions, however, we add a certain degree of regularity. We clarify practical implications of these assumptions in what follows. We note that monotonicity (Assumption~A3) is the most restrictive and crucial assumption of this framework. We also remark that the space of control signals (i.e., temporal pulses~\eqref{eq:pulse}) is rich enough to solve some control problems under Assumptions~{A1 -- A5}. Due to space limitations we refer the reader to~\cite{sootla2016optimalswitching} for detail.

At the first glance, the problem of convergence to $\cB_\varepsilon(x^\bullet)$ (i.e., Problem~1) does not seem to be easier than the problem of convergence to the $\varepsilon$-ball around $x^\bullet$ (e.g., $\{x\in \R^n | \|x-x^\bullet\|_2 \le \varepsilon\}$). As shown in~\cite{sootla2016optimalswitching}, however, Problem~1 can be solved using the following static optimization program under Assumptions~{A1 -- A5}:
\begin{align}
\label{prog:static}\gamma^\ast= \min\limits_{\mu\ge 0, \tau\ge 0}~~   & \ln|r(x, \mu, \tau)| + |\lambda_1| \tau\\
\label{con:time}\text{subject to:}~~&  r(x,\mu,\tau)\le -\varepsilon, \,\, \mu\cdot \tau \le \cE_{\rm max},
\end{align}
where $\cT_{\rm conv}=\frac{1}{|\lambda_1|} (\gamma^\ast-\ln(\varepsilon))$. It can be shown that the function $\ln|r(x,\mu,\tau)| + {|\lambda_1|\tau}$ is decreasing in $\mu$ and $\tau$. Hence the minimum is attained only while one of the constraints is activated. Therefore, we spend all the energy budget $\cE_{\rm max}$ and/or reach the isostable $\cB_\varepsilon$. More specifically, 
if the curves $\{x \in \R^n| s_1(x) = -\varepsilon \}$ and $\mu = \cE_{\rm max}/\tau$ intersect, then the optimal solution lies on this intersection. If these curves do not intersect then we need to solve:
\begin{align}
\label{prog:static2}\gamma^\ast= \min\limits_{\tau\ge 0}~~ \ln\left|r\left(x, \frac{\cE_{\rm max}}{\tau}, \tau\right)\right|  + |\lambda_1| \tau.
\end{align}

The computation of the function $r$ can be performed using the methods to compute $s_1$~\cite{sootla2016optimalswitching}, which may not be numerically cheap for large-scale and/or stiff systems. However, solving an optimal control problem using dynamic programming or maximum principle generally requires (at least) a comparable computational effort. Furthermore, we need to solve a dynamic optimization problem instead of the static one as in~\eqref{prog:static}. 
	
In order to apply this framework to wider class of problems, we briefly comment on the main properties of the solution in~\cite{sootla2016optimalswitching}. Consider the following auxiliary control problem:
\begin{align}
V(z, \mu, \beta) = &\inf\limits_{\begin{smallmatrix} \tau, u \in \cU_\infty([0,\mu]) \end{smallmatrix}}  \tau, \label{prob:opt-escape} \\
\notag \text{ subject to~} &\dot x =f(x, u),\, x(0) = z,\, \\ 
\notag &x(\tau) \in \cC_\beta = \{\tilde x \in \R^n | s_1(\tilde x) = \beta \},
\end{align}
where $\beta\in\R$. Under Assumptions~A1 -- A5, it can be shown that the optimal solution to~\eqref{prob:opt-escape} has the following form:
\begin{gather}
u(x) = \begin{cases}
\mu    &  \text{if } s_1(x) < \beta\\
0      &  \text{if }s_1(x) \ge \beta
\end{cases},\label{pulse:closed-loop-escape}
\end{gather}
The function $r(x,\mu,\tau)$ is related to the function $V(x,\mu,\beta)$, however, this relation is not bijective. The relevant properties of $r$ are listed below.

\begin{prop}[\cite{sootla2016optimalswitching}]\label{thm:r-prop}
	Let the system $\dot x =f(x, u)$ satisfy Assumptions~{A1--A5}, let $s_1(x)<\beta$ and $(x,\mu,\tau)\in\dom(r)$.  \\
	(i) $r$ is a $C^1$ function on $\dom(r)$. Furthermore, $\partial_\tau r(x,\mu,\tau) > \lambda_1 r(x,\mu,\tau)$,   $\partial_\mu r(x,\mu,\tau) > 0$  and $\nabla_x    r(x,\mu,\tau) \gg 0$.\\
	(ii) If $r(x,\mu,\tau)\le 0$, then $\partial_\tau r(x,\mu,\tau)> 0$.\\
	(iii) If $f(x, \nu)\succeq 0$, then $\partial_\tau r(x,\mu,\tau)>0$ for all finite $\tau> 0$, and $\mu > \nu$.
\end{prop}

Additionally, if a solution to~\eqref{prob:opt-escape} exists then $V(x, \mu, \beta) = \tau$  if and only if $r(x,\mu,\tau) = \beta$, provided that (ii) or (iii) holds. If Assumptions~{A4} and~{A5} do not hold, then it can be shown that all the inequalities are not strict, that is $\nabla_x r\succeq 0$, $\partial_\tau r(x,\mu,\tau) \ge \lambda_1 r(x,\mu,\tau)$, $\partial_\mu r(x,\mu,\tau) \ge 0$ in (i), $\partial_\tau r(x,\mu,\tau) \ge 0$ in (ii) and (iii). Furthermore, in this case if a solution to~\eqref{prob:opt-escape} exists then $V(x, \mu, \beta) = \tau$  if and only if $\tau = \argmin_{\xi\ge 0} r(x,\mu,\xi) = \beta$.
\subsection{Convergence Under Parametric Uncertainty}\label{ss:param-unc}

We consider the system~\eqref{sys:f} satisfying assumptions {A1--A5} for every $p\in[p_1,p_2]$. Furthermore, we assume that the system~\eqref{sys:f} is monotone on $\cD\times [p_1, p_2]\times \Rnn$. This is an additional assumption with respect to Assumption~A3, since we require monotonicity of the flow with respect to parameter changes. Let $s_1(x,p)$ denote the eigenfunction of the system~\eqref{sys:f} for a fixed parameter $p$ defined on the basin of attraction $\cB(x^\bullet(p))$ of the exponentially stable equilibrium $x^\bullet(p)$. We first formulate  an auxiliary problem:
\begin{align}
&V(z, \mu, \beta, p) = \inf\limits_{\begin{smallmatrix}\tau, u \in \cU_\infty([0,\mu]) \end{smallmatrix}} \tau, \label{prob:opt-control-param} \\
\notag &\text{ subject to~\eqref{sys:f}},\, x(0) = z,\, x(\tau) \in \cA_\beta,
\end{align}
where $\cA_\beta = \{\tilde x\in \R^n | g(\tilde x) = \beta\}$ and $g\in C^1$ and $\nabla g\gg 0$ on $\dom(g)$. We have the following result, the proof of which can be found in Appendix~\ref{app:proofs}.
\begin{prop}\label{prop:mon-val-fun-gen}
	Let the system~$\dot x =f(x, u)$ satisfy Assumptions~A1--A5 for every parameter $p$, and let  $\nabla g(x) \gg 0$. If $g(z) < \beta$ (respectively, $g(z) \ge \beta$), then $u^0(\cdot) = \mu$ (respectively, $u^0(\cdot) = 0$) is an optimal solution of~\eqref{prob:opt-control-param}. 
\end{prop}
Now provided that we fix the function $g(z)$ for all $p\in[p_1, p_2]$ we have the following result with the proof in Appendix~\ref{app:proofs}.

\begin{prop} \label{prop:opt-control-param}
	Let the system~\eqref{sys:f} satisfy Assumptions~{A1--A5} for parameter values $p_1$, $p_2$ and be monotone on $\cD\times [p_1, p_2]\times \Rnn$. Let $p_1 \preceq p_2$, $y \preceq z$, $\nu \le \mu$, $\max(g(z), g(y)) \le \min(\alpha,\beta)$ then:\\
	(i) $\alpha \le \beta$ implies that $V(z, \mu, \alpha, p_2) \le V(y, \nu, \beta, p_1)$, and the result holds with strict inequalities;\\
	(ii) $V(z, \mu, \alpha, p_2) \ge V(y, \nu, \beta, p_1)$ implies that $\alpha \ge \beta$.
\end{prop}

The main application of this result is control of systems under parametric uncertainty. That is, we assume that the bounds on parameter variations are known (for example, $p\in[p_1,p_2]$), however, the exact values of parameters (and hence the eigenfunctions) are unknown. In this case, we have:
\begin{align*}
V(z, \mu, \beta, p_1) \ge V(z, \mu, \beta,p) \ge V(z, \mu, \beta,p_2),
\end{align*}
for all  $p\in[p_1, p_2]$, if  $g(z) < \beta$.
Hence, given an initial condition, target set, and the maximum allowed control, we can estimate the optimal value function. Note that in this case the function $g$ is not necessarily an eigenfunction for~\eqref{sys:f} with the parameter $p$. But letting $g(z)$ being equal to an eigenfunction $s_1(z, p_1)$ (respectively, $s_1(z, p_2))$ we can estimate the upper bound (respectively, the lower bound) on the value function. In order to do so, we simply compute the function $r$ as in the non-parametric case. 

Estimation of the convergence time to the equilibrium is slightly more delicate. We use an extended definition of the function $r$:
\begin{gather*}
r(x, \mu, \tau, p) = s_1(\phi(\tau,x, p,\mu),p).
\end{gather*} 
The convergence time $\cT$ can be computed in the same manner, provided that the parameter value $p$ is given:
\begin{gather*}
\cT(x, \mu, \tau, p, \varepsilon) = \frac{1} {|\lambda_1(p)|} \ln\left(\frac{|r(x,\mu,\tau, p)|}{\varepsilon}\right).
\end{gather*} 
In the non-parametric case, it was shown that $\partial_\mu r(x, \mu, \tau)>0$ and $\partial_\tau r(x, \mu, \tau)>0$ (if $r(x,\mu,\tau) < 0$) under Assumptions~{A1 -- A5}. Therefore, if we increase $\mu$ or $\tau$ we can predict the changes in $r(x, \mu, \tau)$ and, as a consequence, in $\cT(x, \mu, \tau, \varepsilon)$. In the parameter dependent case, estimating the dependence of the function $r(x, \mu, \tau, p)$ on $p$ is not straightforward. For example, $\partial r(x, \mu, \tau, p)/\partial p$ does not necessarily have the same sign pattern for fixed $x$, $\mu$, and $\tau$, while varying $p$. Furthermore, $\lambda_1(p)$ now also depends on parameters. However, since we are dealing with monotone systems, it is still possible to bound the convergence time provided that the functions $r(\cdot, \cdot, \cdot, p_1)$ and $r(\cdot, \cdot, \cdot, p_2)$ are computed.
\begin{prop}\label{prop:param-iso}
Let the system~\eqref{sys:f} satisfy Assumptions~{A1--A5} for parameter values $p \in[p_1,~ p_2]$ and be monotone on $\cD\times [p_1, p_2]\times \Rnn$.	Let the pair $(\mu,\tau)$ belong to the set
\begin{multline*}
\cS_\sigma(x, p_1, p_2, \varepsilon) = \Bigl\{(\nu,\xi)\in\R^2\Bigl| r(x,\nu,\xi, p_1) \ge -\varepsilon e^{|\lambda_1(p_1)| \sigma},  \\
r(x,\nu,\xi,p_2) \le -\varepsilon e^{|\lambda_1(p_2)| \sigma}\Bigl\},
\end{multline*}
for a positive $\varepsilon$. Then the flow $\phi(t, x, p, \mu h(\cdot,\tau))$ at time $\sigma + \tau$ belongs to the set $\cA_{\rm target} = \{x\in\R^n | z_1 \preceq x \preceq z_2, z_1 \in \partial_-\cB_\varepsilon(x^\bullet(p_1)), z_2 \in \partial_-\cB_\varepsilon(x^\bullet(p_2))\} $.
\end{prop}

The proof in is Appendix~\ref{app:proofs}. We note that $\cS_\sigma(x, p, p, \varepsilon) = \{ (\nu, \xi) \in \R^2\bigl| \cT(x,\nu,\xi,p, \varepsilon) = \sigma \}$, provided that $r$ is negative on this level set. In this case, for $p \in [p_1, p_2]$ we have that  
\begin{multline*}
\cS_\sigma(x, p_1, p_2, \varepsilon) = \bigl\{(\mu,\tau)\in\R^2\Bigl| \nu_1 \le \mu \le \nu_2, \xi_1 \le \tau \le \xi_2 \\
(\nu_1,\xi_1) \in \cS_\sigma(x, p_1, p_1,\varepsilon), (\nu_2,\xi_2) \in \cS_\sigma(x, p_2, p_2, \varepsilon)   \bigl\},
\end{multline*}	
which implies that we can compute the level sets $\cT(x,\mu,\tau, p_1,\varepsilon)$ and $\cT(x,\mu,\tau, p_2,\varepsilon)$ in order to estimate $\cS_\sigma(x, p_1, p_2, \varepsilon)$. Note however, that these levels sets can potentially intersect for large values of $\mu$, which is clearly seen for large values of $\varepsilon$. Therefore, the level sets $\cT(x,\mu,\tau, p,\varepsilon) = \sigma$ do not lie between the level sets $\cT(x,\mu,\tau, p_1,\varepsilon) = \sigma$ and $\cT(x,\mu,\tau, p_2,\varepsilon) = \sigma$ provided that $p\in[p_1,~p_2]$.
Furthermore, only for small $\varepsilon$, the target set $\cA_{\rm target}$ provides a good estimate of the interval $[x^\bullet(p_1), x^\bullet(p_2)]$, which contains $x^\bullet(p)$. The approximation of the convergence time can hence be conservative for large intervals $[p_1, p_2]$ and for large $\mu$. 

We also remark that these results can potentially be extended to a class non-monotone systems $\dot x = f(x, u)$ by using bounding monotone systems approach (cf.~\cite{gouze2000interval}). If there are monotone systems $\dot x = r(x, u)$, $\dot x = g(x,u)$ on $\cD\times \R$ such that $r(x,u) \succeq f(x,u) \succeq g(x,u)$ for all $(x,u) \in \cD\times \R$, then the convergence time for the system $\dot x = f(x, u)$ can potentially be bounded by the convergence time of the systems $\dot x = r(x, u)$, $\dot x = g(x,u)$. 
\begin{figure}[t]
	\centering
	\includegraphics[height = 0.4\columnwidth]{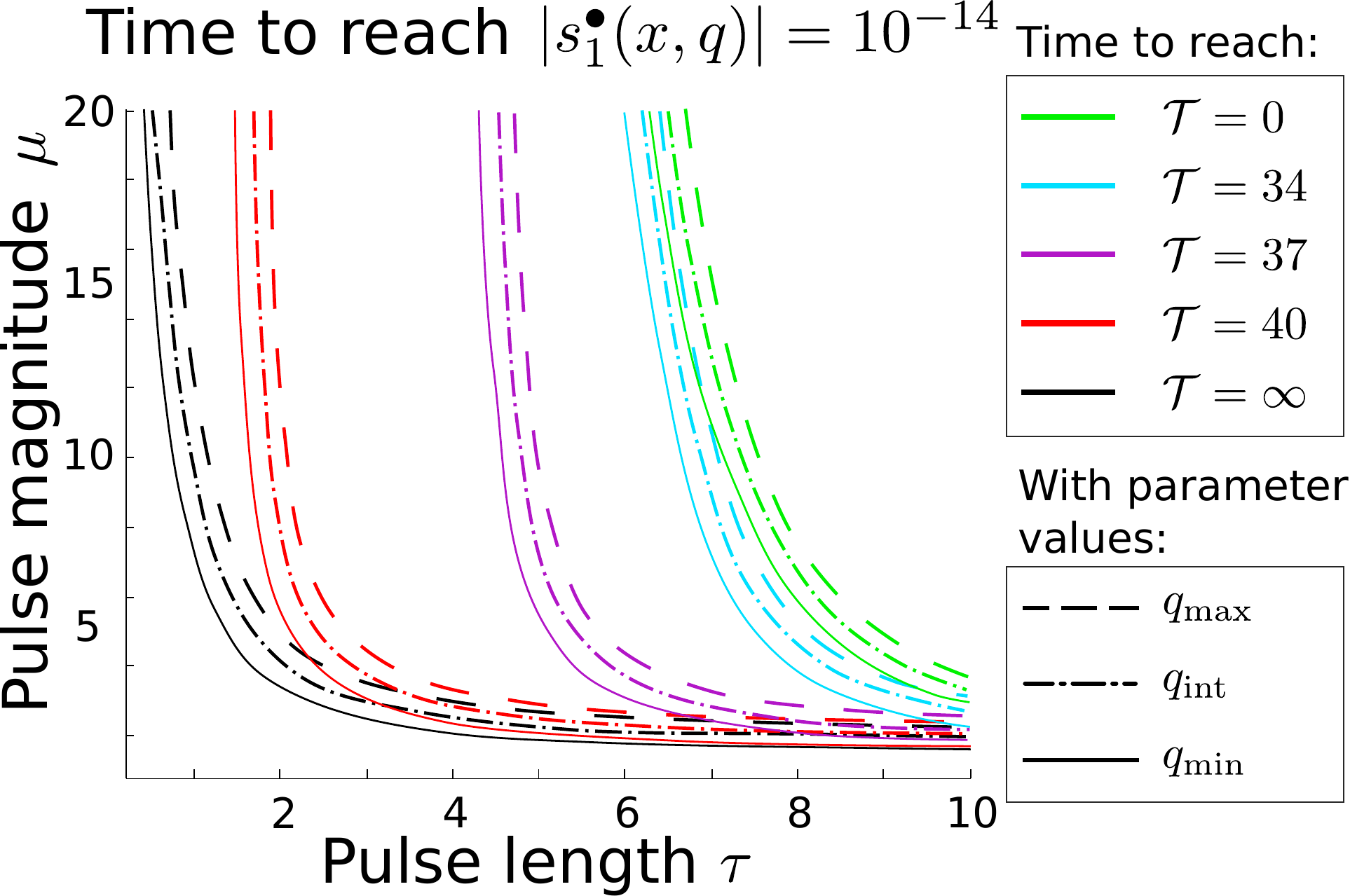}
	\caption{The level sets of $\cT = \frac{1}{|\lambda_1(q)|} \ln \left(\frac{|r(x^\ast(q), \mu, \tau, q)|}{\varepsilon} \right)$ for $q$ equal to $q_{\rm min}$, $q_{\rm int}$, and $q_{\rm max}$, where $\varepsilon = 10^{-14}$.} \label{fig:switching-iso-ts}
\end{figure}

\subsection{A Case Study} 
As a case study we consider a genetic toggle switch system which is met in many applications (cf.~\cite{Gardner00}). We consider the following model with $p_{i j} \ge 0$ for all $i = 1,2$, $j =1,\dots, 4$:
\begin{gather}\label{sys:toggle-2d}
\begin{aligned}
\dot x_1 &= p_{1 1} + \frac{p_{1 2}}{ 1 + x_2^{p_{1 3}}} - p_{1 4} x_1 + u_1, \\
\dot x_2 &= p_{2 1} + \frac{p_{2 2}}{ 1 + x_1^{p_{2 3}}} - p_{2 4} x_2 + u_2.
\end{aligned}
\end{gather}
The states $x_{i}$ represent the concentration of proteins, whose mutual repression is modeled via a rational function. The parameters $p_{1 1}$ and $p_{2 1}$ model the basal synthesis rate of each protein. The parameters $p_{1 4}$ and $p_{2 4}$ are degradation rate constants, $p_{1 2}$, $p_{2 2}$ describe the strength of mutual repression, while $p_{1 3}$, $p_{2 3}$ are called Hill coefficients. We consider the following parameter values:
\begin{gather*}
p = \begin{pmatrix}
q_1 & q_2 & 4 & 1 \\ q_3 & q_4 & 3 & 2
\end{pmatrix},
\end{gather*}
with the set of admissible parameters is equal to $[q_{\rm min}, q_{\rm max}]$, where $q_{\rm min} = \begin{pmatrix}
1.8 & 950 &  1.2 & 1050 
\end{pmatrix}$ and
$q_{\rm max} = \begin{pmatrix}
2.2 & 1050 & 0.7 & 950 
\end{pmatrix}$, and the order $\succeq_q$ is induced by $\cQ =\diag{\begin{pmatrix} 1 & 1 & -1 & -1\end{pmatrix}}\Rnn^4$. We also consider $q_{\rm int} = \begin{pmatrix} 2 & 1000 & 1 & 1000 
\end{pmatrix}$, where $q_{\rm int}\in [q_{\rm min}, q_{\rm max}]$. We can readily check that the model is monotone on $\Rnn^2\times \Rnn^4 \times\Rnn^2$ with respect to $\diag{\begin{pmatrix} 1 & -1 \end{pmatrix}} \Rnn^2\times\cQ \times \diag{\begin{pmatrix} 1 & -1 \end{pmatrix}} \Rnn^2$.  For every $q \in [q_{\rm min}, q_{\rm max}]$ the system is bistable with two exponentially stable equilibria $x^\ast$ and $x^\bullet$.
In this setting, the equilibrium $x^\ast$ has the state $x_2$ ``switched on'' ($x_2$ is much larger than $x_1$), while $x^\bullet$ has the state $x_1$ ``switched on'' ($x_1$ is much larger than $x_2$). The rest of assumptions also hold, which can be verified by direct calculation.

First we illustrate the application of Proposition~\ref{prop:param-iso}. We assume that $u_2 = 0$ and we compute $u_1$ in order to switch from $x^\ast$ to $x^\bullet$.

\emph{Problem 2. Switching Under Parametric Uncertainty.} Consider a system~\eqref{sys:toggle-2d} with $q\in [q_{\rm min}, q_{\rm max}]$. Compute an open-loop policy $u_1=\mu h(\cdot,\tau)$ such that the flow $\phi(t,x^\ast(q), q, \mu h(\cdot,\tau), 0)$ converges to $x^\bullet(q)$ in approximately $\cT+\tau$ time units for any $q\in[q_{\rm min}, q_{\rm max}]$.

In Figure~\ref{fig:switching-iso-ts} we depict the level sets of the function $\cT(\mu, \tau, q, \varepsilon) = \frac{1}{|\lambda_1(q)|} \ln(r(x^\ast(q),\mu,\tau,q)/\varepsilon)$ for parameter values $q_{\rm min}$, $q_{\rm max}$, and $q_{\rm int}$. For $\varepsilon = 10^{-14}$, we can bound the location of the level sets  of $\cT(\mu, \tau, q_{\rm int}, \varepsilon)$ by the level sets of $\cT(\mu, \tau, q_{\rm max}, \varepsilon)$ and $\cT(\mu, \tau, q_{\rm min}, \varepsilon)$ thus solving \emph{Problem 2} under parametric uncertainty. The function $\cT$ appears to change monotonically in $q$, however, for $\varepsilon = 10^{-2}$ the level sets of $\cT$ actually intersect, which rules out monotonicity of $\cT$. Similar observations were made on different sets of parameter values. However, despite this conservatism a set of admissible $\mu$, $\tau$ solving \emph{Problem 2} can still be found.

Next, we consider a slightly more complicated problem and illustrate the application of Proposition~\ref{prop:opt-control-param}.
	
\emph{Problem 3. Event-Based Regulation Around a Saddle Point.} Consider a system~\eqref{sys:toggle-2d}.  Let $x^\dagger(q)$ denote the saddle point, and let $x^\dagger(q)\in[z, y]$. Keep the flow $\phi(t,x ,q, u_1, u_2)$ in $[z, y]$ for any $q\in [q_{\rm min}, q_{\rm max}]$ for all $t\ge T$ for a large enough $T>0$.

\begin{figure}[t]
		\centering
		\includegraphics[height = 0.4\columnwidth]{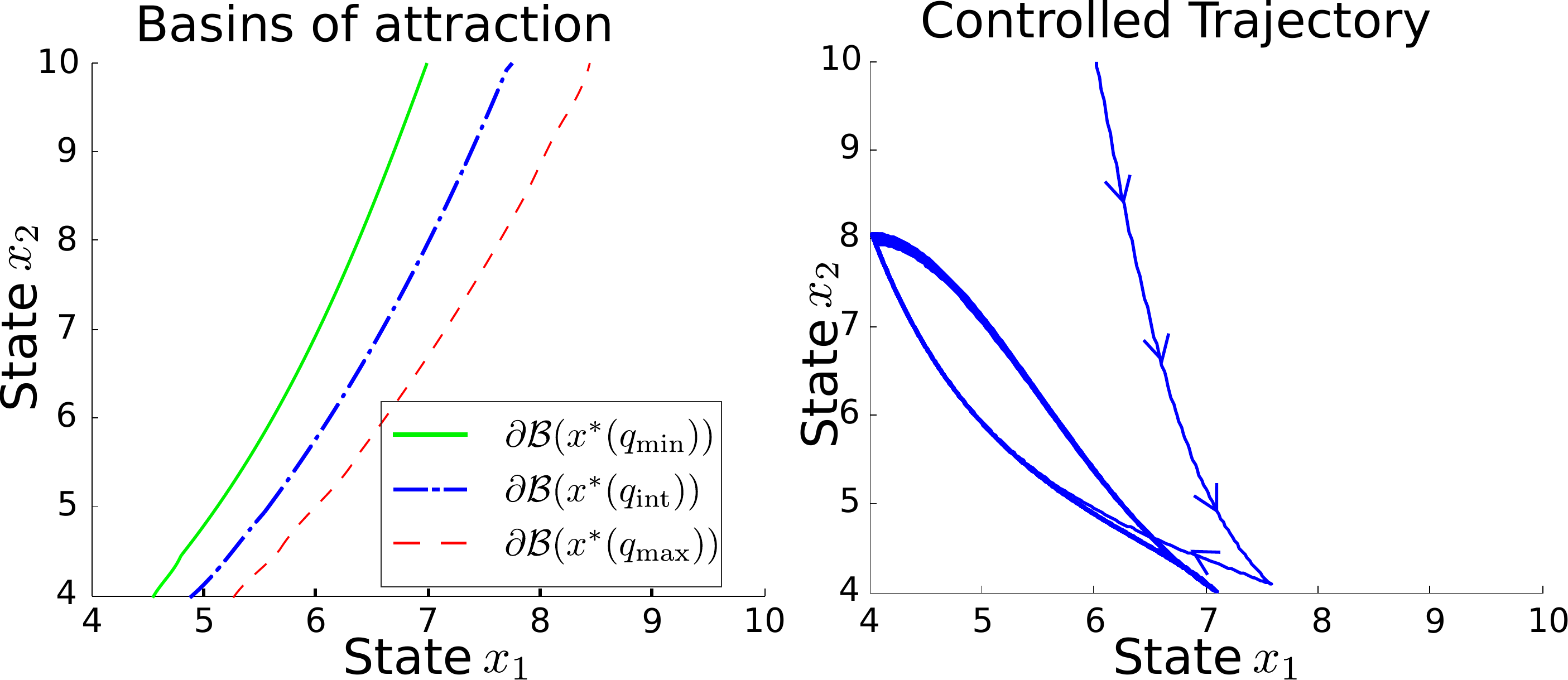}
		\caption{In the left panel, boundaries of basins of attraction of the toggle switch for different parameter values. In the right panel a trajectory controlled by an event-based procedure in order to keep the flow inside the box $[4,10]\times[4, 10]$.} \label{fig:unst-reg}
	\end{figure}

 First we note that $q_{\rm min}\preceq_q q \preceq_q q_{\rm max}$ implies that $\cB(x^\bullet(q_{\rm min}))\subseteq  \cB(x^\bullet(q)) \subseteq \cB(x^\bullet(q_{\rm max}))$, and  $\cB(x^\ast(q_{\rm min})) \supseteq \cB(x^\ast(q)) \supseteq \cB(x^\ast(q_{\rm max}))$ according to results in~\cite{sootla2016basins}, which is illustrated in the left panel of Figure~\ref{fig:unst-reg}. 
 Given the geometry of the systems, we derived an ad-hoc event-based solution as follows.

\emph{Proposed Event-Based Control Scheme:} Let $\delta \ll 1$. If $x = \phi(t, x, p, u_1, 0)\not \in[z, y]$ compute $u_2$ so that $\phi(t, x, p, 0, u_2)$ reaches $\partial_-\cB_{1/\delta}(x^\bullet(q_{\rm min}))$ in minimum-time. If $x = \phi(t, x, p, 0, u_2)\not \in[z, y]$ compute $u_1$ so that $\phi(t, x, p, u_1, 0)$ reaches $\partial_+\cB_{1/\delta}(x^\ast(q_{\rm max}))$ in minimum time. 

Let $y = \begin{bmatrix}
10, & 4
\end{bmatrix}^T$, $z = \begin{bmatrix}
4, & 10
\end{bmatrix}^T$. First, we consider the treatment of the event $\phi(t, x, p, u_1, 0)\not \in[z, y]$. We pick a finite number of points $\hat x_i$ on the boundary of $[z, y]$, and compute the magnitudes for the control signals $u_2 = \mu h(\cdot, \xi_{\rm upper})$ with $\xi_{\rm upper} = 10$ steering the flow from $\hat x_i$ to $\partial_-\cB_{1/\delta}(x^\bullet(q_{\rm min}))$. Note that we will reach $\partial_-\cB_{1/\delta}(x^\bullet(q))$ in time $\tau < 10$ due to Proposition~\ref{prop:opt-control-param}. For a small enough $\delta$ we also have that  $\cB_{1/\delta}(x^\bullet(q_{\rm min})) \subseteq \cB_{1/\delta}(x^\bullet(q_{\rm max}))$ around the saddle points. Therefore we can also estimate the lower bound on $\tau$ by computing $\xi_{\rm lower}$ such that $r(\hat x_i, \mu, \xi_{\rm lower},q_{\rm min}) = -1/\delta$. However, these estimates are not accurate for a large $\delta$ since isostables do not change monotonically under parameter variation (see~\cite{sootla2016basins}). When we detect that $x=\phi(t, x, p, u_1, 0)\not \in [z,y]$ we find the closest to $x$ points $\hat x_i$, and steer the flow from $\hat x_i$ with $u_2 = \mu h(\cdot, \xi_{\rm upper})$. The event $x = \phi(t, x, p, 0, u_2)\not \in[z, y]$ is treated in a similar way. In this case, we use $u_1$ and aim to reach the isostable $\partial_+\cB_{1/\delta}(x^\ast(q_{\rm max}))$ in $10$ time units. 

A resulting trajectory for $q_{\rm int}$ is depicted in Figure~\ref{fig:unst-reg}. We note the flow cannot be stabilized by a single control signal in the set $[z,~y]$, since for all constant control signal values only saddle points are contained in $[z,~y]$. Therefore, limiting the flow to a set around an unstable equilibrium is the only available solution.

\section{Conclusion}
In this technical note, we studied pulse-based control approach to the convergence problem. We extended this approach to the convergence problem for monotone systems under parametric uncertainty, where we showed how to estimate the convergence time to the target equilibrium. We illustrated our results on numerical examples, and we showed how our algorithm can be applied to different control problems. 

\bibliography{optimal-control-tran-final.bbl}
\appendices
\section{Proofs}
\label{app:proofs}
	\begin{proof-of}{Proposition~\ref{prop:mon-val-fun-gen}} 
This proof is along the lines of the proof of Theorem~8 in~\cite{sootla2016optimalswitching}, therefore we only provide a sketch. Let $s_1(z) <\beta$, while the case $s_1(z) \ge \beta$ is shown similarly. Due to the premise of Proposition~\ref{prop:mon-val-fun-gen}, $s_1(z_1) \ge s_1(z_2)$ for all $z_1 \succeq z_2$. Using Assumptions A3 -- A5, it can be shown that $\phi(t, x, \mu) \succeq \phi(t, x, u)$ for all $t>0$ and any $u(t)$ such that  $\mu \ge u(t)$ for all $t>0$, therefore $g(\phi(\tau, x, \mu)) \ge g(\phi(\tau, x, u))$. Therefore, by using the control signal $u^0(t) = \mu$ for all $t$ we reach the level set $\cA_\beta$ at least as fast as we any other control signal $u(t) \le \mu$ for all $t\ge 0$. Since we consider only $u\in \cU_{\infty}([0,\mu])$ the control signal $u^0(\cdot) = \mu$ is an optimal solution to the considered problem.	
\end{proof-of}
\begin{proof-of}{Proposition~\ref{prop:opt-control-param}}
	(i) Let $\cC_\alpha^- = \{x | g(x) \ge \alpha\}$, $\cC_\beta^- = \{x | g(x) \ge \beta\}$. Due to this definition we have that $\cC_\alpha^- \supseteq \cC_\beta^-$ if and only if $\alpha\le\beta$. By the premise $g(z) < \alpha$, $g(y) < \beta$, hence we can view the problem as reaching the sets $\cC_\alpha^-$, $\cC_\beta^-$.  
	
	We proceed by showing that 
	\begin{gather}
	V(y, \nu, \beta, p_1) \ge V(z, \mu, \beta, p_2), \label{ineq2:prop12}
	\end{gather}
	for all $z\succeq y$, $\mu\ge \nu$, $p_2 \succeq p_1$. 	Due to monotonicity we have that $\phi(t, z, \mu, p_2) \succeq \phi(t, y, \nu, p_1)$ for all $t\ge0$, $\mu\ge \nu$, $z\succeq y$, $p_2 \succeq p_1$, which entails that $g(\phi(t, z, \mu, p_2)) \ge g(\phi(t, y, \nu, p_1))$. Hence 	$g(\phi(\xi, z, \mu, p_2)) \ge g(\phi(\xi, y, \nu, p_1)) = \beta$, which implies that the flow $\phi(t, z, \mu, p_2)$ has entered the set $\cC_\beta^-$ at some time $t \le \xi$, so that $V(z, \mu, \beta, p_2) \le V(y, \nu, \beta, p_1)$.  
	
	Additionally since $\cC_\alpha^-\supseteq\cC_\beta^-$ we have that 
	\begin{equation}\label{ineq1:prop12}
	V(z, \mu, \beta, p_2) \ge V(z, \mu, \alpha, p_2), \text{ for } \beta\ge \alpha,
	\end{equation}
	since it takes more time to reach  $\cC_\beta^-$ with the same optimal control signal. Now using~\eqref{ineq2:prop12} we get
	$V(y, \nu, \beta, p_1)  \ge V(z, \mu, \beta, p_2) \ge V(z, \mu, \alpha, p_2)$ and complete the proof. 
	
	If $\alpha < \beta$, then using the same arguments we can obtain the following chain of inequalities $V(y, \nu, \beta, p_1)  \ge V(z, \mu, \beta, p_2) > V(z, \mu, \alpha, p_2)$ for $\mu \ge \nu$, $z\succeq y$, $p_2\succeq p_1$
	
	(ii) We prove this part by contradiction. Let $V(z, \mu, \alpha, p_2) \ge V(y, \nu, \beta, p_1)$ and $\alpha < \beta$ for some $\mu \ge \nu$, $z\succeq y$ . According to the point (i), we have that $V(z, \mu, \alpha, p_2) < V(y, \nu, \beta, p_1)$, which contradicts our assumption. 
\end{proof-of}

\begin{proof-of}{Proposition~\ref{prop:param-iso}} 	
	Let $(\mu,\tau)$ belong to $\cS_\sigma(x, p_1, p_2, \varepsilon)$, and let  $s_1(\phi(\tau, x, p_1, \mu), p_1) \ge - \varepsilon e^{ -\lambda_1(p_1)\cT}$  and $s_1(\phi(\tau, x, p_2, \mu), p_2) \le -\varepsilon e^{ -\lambda_1(p_2)\sigma}$. Due to monotonicity for all $p\in [p_1, p_2]$ we have
	\begin{multline*}
	\phi(\tau + \sigma, x, p_1, \mu h(\cdot, \tau)) \preceq  \phi(\tau + \sigma, x, p, \mu h(\cdot, \tau)) \preceq \\ \phi(\tau + \sigma, x, p_2, \mu h(\cdot, \tau))\,\, . 
	\end{multline*}
	Now, since $s_1(\cdot, p)$ is increasing and $s_1(\phi(t,x, p, 0), p) = s_1(x, p) e^{\lambda_1(p) t}$ for any $p\in[p_1,~p_2]$, we have 
	\begin{multline}
	s_1(\phi(\tau + \sigma, x, p, \mu h(\cdot, \tau)), p_1) \ge  \\
	s_1(\phi(\tau + \sigma, x, p_1, \mu h(\cdot, \tau)), p_1) =  \\
	s_1(\phi(\tau , x, p_1, \mu ), p_1) e^{\lambda_1(p_1) \sigma} \ge -\varepsilon. \label{eq:prop8a}
	\end{multline}	
	Similarly, we can obtain, the following chain of inequalities:
	\begin{multline}
	s_1(\phi(\tau + \sigma, x, p, \mu h(\cdot, \tau)), p_2) \le \\
	s_1(\phi(\tau + \sigma, x, p_2, \mu h(\cdot, \tau)), p_2)  =\\
	 s_1(\phi(\tau , x, p_2, \mu), p_2)  e^{\lambda_1(p_2)\sigma} \le -\varepsilon.\label{eq:prop8b}
	\end{multline}
	By noticing that the inequalities~\eqref{eq:prop8a},~\eqref{eq:prop8b}  are the definition of the set $\cA_{\rm target}$, we complete the proof.
\end{proof-of}
\end{document}